\documentclass[%
aip,
amsmath,amssymb,
reprint,%
]{revtex4-1}
\usepackage{graphicx}
\usepackage{dcolumn}
\usepackage{bm}
\usepackage{color} 
\usepackage{physics}
\usepackage{hyperref}
\usepackage{subfigure}
\hypersetup{
    colorlinks = true,
    linkcolor = blue,
    citecolor = blue,
    anchorcolor = blue,
    urlcolor = blue
    }
\newcommand{\red}[1]{{#1}}

\def\lesssim{\ \raise.3ex\hbox{$<$}\kern-0.8em\lower.7ex\hbox{$\sim$}\ }
\def\gesim{\ \raise.3ex\hbox{$>$}\kern-0.8em\lower.7ex\hbox{$\sim$}\ }
\def\up{\uparrow}
\def\dwn{\downarrow}

\newcommand \beq{\begin{eqnarray}}
\newcommand \eeq{\end{eqnarray}}
\newcommand{\sign}[1]{\mathrm{sgn}{#1}}

\usepackage{fixmath}
\usepackage[normalem]{ulem}

\usepackage{comment}

\begin{document}

\title{Dominant Andreev Reflection through Nonlinear Radio-Frequency Transport}

\author{Tingyu Zhang}
\affiliation{Department of Physics, School of Science, The University of Tokyo, Tokyo 113-0033, Japan}
\author{Hiroyuki Tajima}
\email{hiroyuki.tajima@tnp.phys.s.u-tokyo.ac.jp}
\affiliation{Department of Physics, School of Science, The University of Tokyo, Tokyo 113-0033, Japan}
\author{Yuta Sekino}
\affiliation{RIKEN Cluster for Pioneering Research (CPR), Astrophysical Big Bang Laboratory (ABBL), Wako, Saitama, 351-0198 Japan}
\affiliation{Interdisciplinary Theoretical and Mathematical Sciences Program (iTHEMS), RIKEN, Wako, Saitama 351-0198, Japan}
\author{Shun Uchino}
\affiliation{Advanced Science Research Center, Japan Atomic Energy Agency, Tokai, Ibaraki 319-1195, Japan}
\author{Haozhao Liang}
\affiliation{Department of Physics, School of Science, The University of Tokyo, Tokyo 113-0033, Japan}
\affiliation{Interdisciplinary Theoretical and Mathematical Sciences Program (iTHEMS), RIKEN, Wako, Saitama 351-0198, Japan}

\begin{abstract}
\section*{Abstract}
 \red{It is found that Andreev reflection provides a deterministic teleportation process at an ideal normal-superconductor interface, making it behave like an information mirror. However, it is challenging to control the Andreev reflection in a spatially-separated junction due to the mode mixing at the interface.} We theoretically propose the laser-induced Andreev reflection between two-component Fermi superfluid and normal states without mode mixing via spatially-uniform Rabi couplings. By analyzing the tunneling current up to the fourth order, we find that the Andreev  current exhibits unconventional non-Ohmic transport at zero temperature. The Andreev current gives the only contribution in the synthetic junction system at zero detunings regardless of the ratio of the chemical potential bias to the superfluid gap, which is in sharp contrast to that in conventional junctions. \red{Our result may give a potential impact on theoretical and experimental study of quantum many-body phenomena}, and also pave a way for understanding the black hole information paradox through the Andreev reflection as a quantum-information mirror.
\end{abstract}

\maketitle

\section*{Introduction}
The study of transport phenomena in ultracold atomic systems can greatly improve our understanding of quantum many-body problems owing to controllability of microscopic parameters.  
By using Feshbach resonances~\cite{RevModPhys.82.1225}, 
one can tune the interparticle scattering length, allowing to scan quantum many-body systems from 
the weakly-interacting to strongly-correlated regimes. 
This technique has successfully been applied to study ultracold Fermi gases in terms of crossover between
the Bardeen-Cooper-Schrieffer (BCS) and Bose-Einstein Condensation (BEC) regimes~\cite{PhysRevLett.92.040403,PhysRevLett.92.120401,BCS-BEC}.
More recently, a variety of experiments with such Fermi gases have been done to observe various quantum transport phenomena including the direct current transport of bulk and mesoscopic systems~\cite{krinner2017,enss2019}.

One of these topics of current interest is the Andreev reflection~\cite{osti_4071988}, originally introduced by Andreev 
to explain the anomalous resistance of heat flow through a normal state-superconducting (N-S) interface. 
The Andreev process involves a conversion between a particle and a hole-like mode as well as creation or annihilation of a condensed pair in the BCS ground state, and exhibits unique characteristics different from conventional tunnelings~\cite{tinkham2004,asano}.
In electron systems, the Andreev reflection has also attracted attention 
in terms of quantum tunneling phenomena such as the proximity effect~\cite{Pannetier2000,Klapwijk2004}
and the Josephson effect~\cite{asano}.
In addition, the presence of the Andreev reflection has  been reported in charge neutral systems such as liquid Helium~\cite{PhysRevLett.70.1846} and ultracold Fermi gas~\cite{doi:10.1126/science.aac9584}.

Recently, it was pointed out that the Andreev reflection can be regarded 
as an analogue of Hawking radiation at a black hole event horizon~\cite{PhysRevD.96.124011,PhysRevD.98.124043,PhysRevD.102.064028}. 
By assuming a momentum-conserved tunneling, the Andreev reflection can 
provide an information-mirroring process which is similar to the black hole evaporation as Hayden and Preskill's proposal~\cite{Hayden_2007,Lloyd2014UnitarityOB}. 
In spite of the interesting connection, in reality it is challenging to control the Andreev reflection in the spatially-separated junction  
in which mode mixing at the interface occurs. 
Therefore, a specific system to experimentally simulate the information-mirror process is  still lacking.

\begin{figure}[tp]
    \centering
    \includegraphics[width=8.6cm]{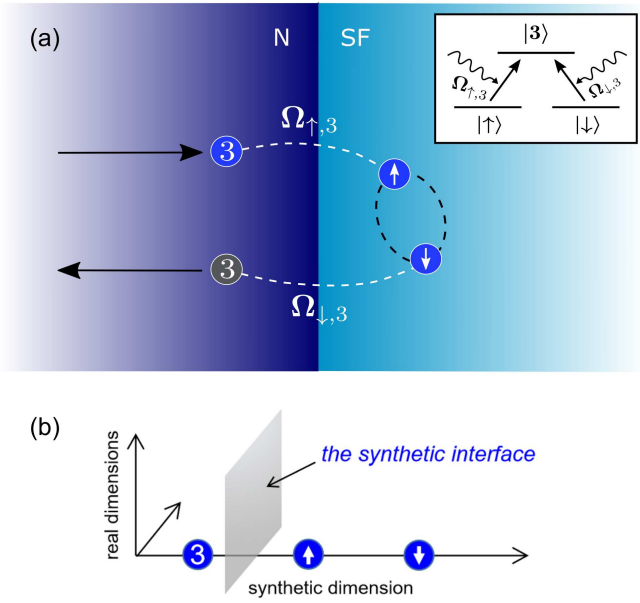}
     \caption{{\bf Schematic of Andreev reflection through an synthetic N-S interface.} (a) An incident particle (hole) is retroreflected as a hole (particle) into the same normal-state reservoir. The state $\ket{3}$ in the normal side interacts with states $\ket{\up}$ and $\ket{\dwn}$ in the superfluid side via the Rabi coupling $\Omega_{\up,3}$ and $\Omega_{\dwn,3}$, respectively. \red{The N and SF represents the normal and superfluid phase.} The energy level diagram for the laser-induced state transition is shown in the inset. (b) The synthetic interface at an extra synthetic dimension.}
     \label{schematic}
\end{figure}

In this work, we  propose a system provoking the momentum-conserved Andreev reflection without the mode mixing by applying multiple radio-frequency (rf) laser fields (see Fig.~\ref{schematic}(a)).
While the rf spectroscopy in ultracold atomic gases~\cite{PhysRevLett.122.203402} has been harnessed 
to extract the quasiparticle excitation, 
we consider double rf laser fields which transfer two hyperfine states $\ket{\up}$ and $\ket{\dwn}$ in the BCS superfluid phase to the third hyperfine state $\ket{3}$ in the normal phase,
to realize an effective N-S interface 
on the internal space~\cite{doi:10.1126/science.1100782,doi:10.1126/science.1100818}. To illustrate this synthetic interface, beyond real dimensions we propose an extra synthetic dimension, where each site denotes an internal state of atoms (Fig.~\ref{schematic}(b)). The synthetic interface separates the normal state $\ket{3}$ from the superfluid state $\ket{\up}$ and $\ket{\dwn}$, and the analogy between real spaces and internal spaces is valid regardless of nonlinear transport processes~\cite{science.aaa8736,stuhl2015visualizing}.

Following the idea above and using the Schwinger-Keldysh formalism, we study the laser-induced tunneling current between the superfluid- and normal-state reservoirs driven by the Rabi couplings $\Omega_{\up,3}$ and $\Omega_{\dwn,3}$. 
By analyzing the current up to the fourth order in $\Omega_{\up,3}$ and $\Omega_{\dwn,3}$,
we find that the Andreev reflection appearing at the nonlinear response regime is the only transport process in the junction system
at zero detunings.
Contrary to the conventional wisdom in the N-S systems~\cite{PhysRevB.66.165413,PhysRevB.25.4515,PhysRevB.54.7366,PhysRevResearch.2.023340},
the Andreev current is not suppressed in the supergap regime, where the chemical potential bias between the normal and superfluid reservoirs is greater than the superfluid gap.
Moreover, the momentum-conserved Andreev current exhibits 
a non-Ohmic transport at zero temperature.
Below, we take $k_{\rm B}=\hbar=1$ and the system volume is taken to be unity.

\section*{Results}
\subsection*{Model}
The Hamiltonian of the normal-state reservoir $\ket{3}$ with the energy level $\omega_3$ is given by 
$H_3=\sum_{{\bf{p}}}(\varepsilon_{{\bf{p}}}+\omega_3)c^\dagger_{{\bf{p}},3}c_{{\bf{p}},3}$ with $\varepsilon_{{\bf{p}}}=p^2/(2m)$, and $c^\dagger_{{\bf{p}},3}$ ($c_{{\bf{p}},3}$) creates (annihilates)
a fermion in state $\ket{3}$ with momentum $\bf{p}$. 
The Hamiltonian of the superfluid-state reservoir is taken as 
\begin{equation}
    \begin{aligned}
    H_{\rm SF}=&\sum_{{\bf{k}},\sigma}(\varepsilon_{\bf{k}}+\omega_\sigma)d^\dagger_{{\bf{k}},\sigma}
    d_{{\bf{k}},\sigma} \\
    &-g\sum_{{\bf{k}},{\bf{k}}',{\bf{P}}}
    d_{{\bf{k}}+\frac{{\bf{P}}}{2},\up}^\dagger
    d_{-{\bf{k}}+\frac{{\bf{P}}}{2},\dwn}^\dagger
    d_{-{\bf{k}}'+\frac{{\bf{P}}}{2},\dwn}
    d_{{\bf{k}}'+\frac{{\bf{P}}}{2},\up}, 
    \end{aligned}
\end{equation}
where $d^\dagger_{{\bf{k}},\sigma}$ and $d_{{\bf{k}},\sigma}$ are respectively the creation and annihilation operators for fermions in states $\ket{\sigma=\up,\dwn}$ with momentum $\bf{k}$ and energy level $\omega_\sigma$. 
Here, $g$ is the strength of attractive interaction in the superfluid reservoir~\cite{PhysRev.108.1175}.  
We then introduce Rabi couplings to induce a particle transfer between the reservoirs~\cite{PhysRevA.72.011602,PhysRevLett.85.487,PhysRevA.72.013601,PhysRevA.82.033629}.
Typically, the wavelengths of rf fields are large compared to the size of the atomic gas and 
the spatial dependence of Rabi couplings are ignorable.
Thus, the corresponding Rabi coupling term can be expressed as 
\begin{equation}\label{Ht}
    \begin{aligned}
    H_{\rm t}&=\sum_{{\bf{k}},\sigma}\left(e^{-i\omega_{{\rm L},\sigma}t}\Omega_{\sigma,3}d_{{\bf{k}},\sigma}^\dagger c_{{\bf{k}}, 3}+{\rm H.c.}\right),
    \end{aligned}
\end{equation}
where $\omega_{{\rm L},\sigma}$ is the laser frequency. 
We note that $H_{\rm t}$ retains the momentum conservation.
The total Hamiltonian of the system is thus $H=H_3+H_{\rm SF}+H_{\rm t}$.  
The particle current operator between two reservoirs is defined as 
\begin{equation}
    \begin{aligned}
        \hat{I}=&-\Dot{N}_3=i\left[N_3, H_{\rm t}\right]\\
         =&-i\sum_{{\bf{k}},\sigma}e^{-i\omega_{{\rm L},\sigma}t}\Omega_{\sigma,3}d^\dagger_{{{\bf{k}}},\sigma}
        c_{{\bf{k}},3}+{\rm H.c.},
    \end{aligned}
\end{equation}
where $N_3=\sum_{{\bf{p}}}c^\dagger_{{\bf{p}},3}c_{{\bf{p}},3}$ is the particle number operator of the normal-state reservoir. 
Notice that the current expression above corresponds to the tunneling current expression between the spatially-separated reservoirs~\cite{PhysRevB.54.7366} 
except for the presence or absence of the momentum conservation. 
In what follows, we consider the zero detunings as $\omega_3-\omega_{\up}-\omega_{{\rm L},\up}
=\omega_3-\omega_{\dwn}-\omega_{{\rm L},\dwn}=0$, where the usual quasiparticle current is suppressed~\cite{PhysRevA.64.033609}.

\subsection*{Laser-induced tunneling current.}
To study the tunneling current between the reservoirs, 
Schwinger-Keldysh Green's function formalism~\cite{Schwinger,Keldysh} is applied with the operators evolving with Hamiltonian $H_0=H_3+H_{\rm SF}$. 
After performing the perturbative expansion of the tunneling current with respect to $H_{\rm t}$, we evaluate the correlation functions in each reservoir with thermal equilibrium under the grand-canonical Hamiltonian $K_0=H_0-\mu_3N_3-\mu_{\rm S}N_{\rm S}$.
Here $\mu_3$ and $\mu_{\rm S}$ are the chemical potentials of particles in normal state $\ket{3}$ and superfluid states, respectively, and $N_{\rm S}=N_{\uparrow}+N_{\downarrow}$ is the particle number operator of the superfluid reservoir ($N_{\sigma}$ is the spin-resolved one).
In the following, $a^{(H_0)}(t)$ and $a^{(K_0)}(t)$ denote operator $a$ in the Heisenberg pictures of $H_0$ and $K_0$, respectively.
Using the relations $d^{\dagger(H_0)}_{{\bf{k}},\sigma}(t)=e^{i\mu_{\rm S} t}d^{\dagger (K_0)}_{{\bf{k}},\sigma}(t)$ and $c^{\dagger(H_0)}_{{\bf{k}},3}(t)=e^{i\mu_3 t}c^{\dagger(K_0)}_{{\bf{k}},3}(t)$,
we obtain the expectation value of the current $I\equiv \langle \hat{I}(t,t)\rangle$, where
\begin{equation}\label{Keldysh}
    \begin{aligned}
        \big\langle \hat{I}(t,t')\big\rangle&=-i\sum_{{\bf{k}},\sigma}\sum_{n=0}^{\infty}\frac{(-i)^n}{n!}\int_C dt_1\cdots\int_Cdt_n\\
        &\Omega_{\sigma,3}\big\langle\mathrm{T}_C
        e^{-i(\mu_3 t'-\mu_{\rm S} t)}d^{\dagger(K_0)}_{{\bf{k}},\sigma}(t)
        c^{(K_0)}_{{\bf{k}},3}(t')\\
        &H_{\rm t}(t_1)\cdots H_{\rm t}(t_n)\big\rangle+{\rm H.c.}
    \end{aligned}
\end{equation}
Here, we introduced 
\begin{equation}
    H_{\rm t}(t)=e^{-i\Delta\mu t}\sum_{{\bf{k}},\sigma}\Omega_{\sigma,3}d^{\dagger(K_0)}_{{\bf{k}},\sigma}(t) c^{(K_0)}_{{\bf{k}},3}(t)+{\rm H.c.},
\end{equation}
and $\Delta\mu=\mu_3-\mu_{\rm S}$ denotes the chemical potential bias between two reservoirs. The integral in Eq.~(\ref{Keldysh}) is taken along the Keldysh contour $C$ 
and $\mathrm{T}_C$ is the contour ordering product operator. 
By using the Langreth rules~\cite{stefanucci_van}, we can change the contour of integral into the real time axis, from $t=-\infty$ to $t=+\infty$, and write each perturbation in terms of Green's functions. 
In addition, we introduce the $2\times2$ Nambu representation in which Green's functions adopt the form
\begin{equation}
    i\hat{G}_{d(c)}({\bf{k}},t-t')=\big\langle\mathrm{T}_C A_{d(c)}(t) A^\dagger_{d(c)}(t')\big\rangle,
\end{equation}
with vectors $A_{d}(t)=(d_{{\bf{k}},\up}(t),d_{-{\bf{k}},\dwn}^{\dagger}(t))^T$ and $A_{c}(t)=(c_{{\bf{k}},3}(t),c_{-{\bf{k}},3}^{\dagger}(t))^T$.
\begin{figure}[tp]
    \includegraphics[width=8cm]{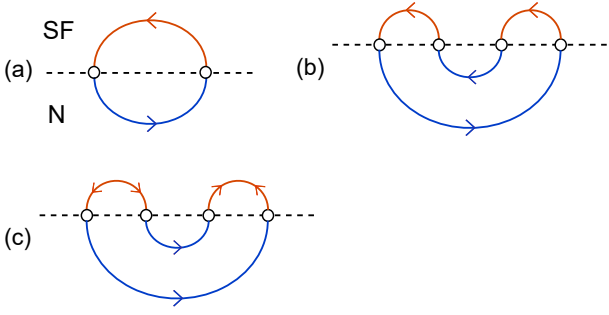}
    \caption{{\bf Diagrammatic representation of tunneling processes between two reservoirs.} The blue (red) lines and the circles represent Green's functions in the normal (superfluid) side and the Rabi couplings 
    $\Omega_{\sigma,3}$.
    (a) Lowest-order diagram, representing the normal single-particle tunneling. (b) Next-to-leading-order diagram, involving the nonlinear quasiparticle tunneling process. (c) Next-to-leading-order diagram, involving the Andreev reflection.}\label{Feynman}
\end{figure}
We note that $\hat{G}_c({\bf{k}},t-t')$ has no off-diagonal elements, while in the superfluid states
the anomalous Green's functions $G_{d,12}$ and $G_{d,21}$ are generated due to the nonzero value of gap parameter~\cite{Fetter}. 
After these manipulations, the leading-order contribution ($n=1$) in frequency representation is obtained as  
\begin{equation}\label{I(1)}
    \begin{aligned}        I^{(1)}=-2\sum_{{\bf{k}},\sigma}&\int\frac{d\omega}{2\pi}\Omega^2_{\sigma,3}
        \operatorname{Re}\Big[G^{{\rm ret.}*}_{d,11}({\bf{k}},\omega)G^<_{c,11}
        ({\bf{k}},\omega-\Delta\mu)\\
        &+G^<_{d,11}({\bf{k}},\omega)G^{\rm ret.}_{c,11}
        ({\bf{k}},\omega-\Delta\mu)\Big],
    \end{aligned}
\end{equation}
with the lesser Green's functions $G^<$ and retarded Green's functions $G^{\rm ret.}$.
By using the relation $G^{<}=-2i\operatorname{Im}[G^{\rm ret.}]f(\omega)$, where $f(\omega)=1/(e^{\omega/T}+1)$ is the Fermi distribution function, one can rewrite Eq.~(\ref{I(1)}) as  
\begin{equation}\label{I(1)1}
    \begin{aligned}
        I^{(1)}=&~ 4\sum_{{\bf{k}}}\int\frac{d\omega}{2\pi}(\Omega^2_{\uparrow,3}+
        \Omega^2_{\downarrow,3})\operatorname{Im}G_{d,11}({\bf{k}},\omega)\\
        &\operatorname{Im}G_{c,11}({\bf{k}},\omega-\Delta\mu)
        \left[f(\omega-\Delta\mu)-f(\omega)\right],
    \end{aligned}
\end{equation}
where, for brevity, we hereafter omit the superscript for all retarded Green's functions.  
Such a current corresponds to the lowest-order single-particle tunneling between the reservoirs as shown in Fig.~\ref{Feynman}(a). 
We note that the above lowest-order analysis corresponds to the linear response theory conventionally adopted in
rf spectroscopy~\cite{PhysRevLett.85.487,PhysRevA.64.033609,PhysRevA.72.013601}.
As we illustrate below, however, the lowest-order analysis is insufficient to discuss the nonzero current between two reservoirs.

\subsection*{Andreev reflection in nonlinear rf current.}
We are now in a position to evaluate the particle current up to the next-to-leading order ($n=3$).
The two coupling constants $\Omega_{\up,3}$ and $\Omega_{\dwn,3}$ give rise to the contractions like $\langle d^\dagger_{-{\bf{k}},\dwn} d^\dagger_{{\bf{k}},\up} \rangle$ and $\langle d_{{\bf{k}},\up} d_{{\bf{k}},\dwn} \rangle$, which do not vanish in the presence of the superfluid.
As a result, such contractions cause tunnelings with pair degrees of freedom including the Andreev reflection.
The total current up to this order is obtained as $I=I^{(1)}+I^{(3)}_1+I^{(3)}_2+I_{\rm A}$, where 
\begin{widetext}
\begin{equation}\label{I1(3)}
    I_1^{(3)}=4\sum_{{\bf{k}},\sigma,\sigma'}\Omega^2_{\sigma,3}\Omega^2_{\sigma'3}
    \int\frac{d\omega}{2\pi}\left[\operatorname{Im}G_{d,11}({\bf{k}},\omega)
    \right]^2\left[\operatorname{Im}G_{c,11}({\bf{k}},\omega-\Delta\mu)\right]^2
    \left[f(\omega-\Delta\mu)-f(\omega)\right],
\end{equation}
\begin{equation}\label{I2(3)}
    I^{(3)}_2=16\Omega^2_{\uparrow,3}\Omega^2_{\downarrow,3}\sum_{{\bf{k}}}\int\frac{d\omega}{2\pi}[\operatorname{Im}G_{d,12}({\bf{k}},\omega)]^2
    \operatorname{Im}G_{c,11}({\bf{k}},\omega-\Delta\mu)\operatorname{Im}G_{c,22}({\bf{k}},\omega+\Delta\mu)
    \big[f(\omega)-f(\omega-\Delta\mu)\big],
\end{equation}
\begin{equation}\label{IA}
    I_{\rm A} = 8\Omega^2_{\uparrow,3}\Omega^2_{\downarrow,3} \sum_{{\bf{k}}}\int\frac{d\omega}{2\pi}|G_{d,12}({\bf{k}},\omega)|^2\operatorname{Im}G_{c,11}({\bf{k}},\omega-\Delta\mu)\operatorname{Im}G_{c,22}({\bf{k}},\omega+\Delta\mu)
    \big[f(\omega-\Delta\mu)-f(\omega+\Delta\mu)\big].
\end{equation}
\end{widetext}
Here $I^{(3)}_1$ is the current corresponding to the nonlinear quasiparticle tunneling between the reservoirs shown in Fig.~\ref{Feynman}(b). On the other hand, $I_2^{(3)}$ and $I_{\rm A}$ originate from the processes represented by Fig.~\ref{Feynman}(c), while the former corresponds to a transfer of a single particle (hole) in normal side with creation or annihilation of pairs in superfluid side as an intermediate state and the latter arises from the Andreev reflection. To see the detailed properties of each contribution,
we take the standard forms of Green's functions in the normal-state reservoir, in which case the imaginary parts are given by $\operatorname{Im}G_{c,11}({\bf{k}},\omega)=\operatorname{Im}G_{c,22}({\bf{k}},-\omega)=-\pi\delta(\omega-\xi_{{\bf{k}},3})$.
For the superfluid reservoir, we take the mean-field form of Green's functions, where the gap parameter $\Delta_{\rm S}=g\sum_{{\bf{k}}'}\langle d_{-{\bf{k}}',\dwn}d_{{\bf{k}}',\up}\rangle$ arises and the imaginary parts read
$\operatorname{Im}G_{d,11}({\bf{k}},\omega)=-\pi[u_{{\bf{k}}}^2\delta(\omega-E_{{\bf{k}}}) +v^2_{{\bf{k}}}\delta(\omega+E_{{\bf{k}}})]$, and $\operatorname{Im}G_{d,12}({\bf{k}},\omega)=\operatorname{Im}G_{d,21}({\bf{k}},\omega)=-u_{{\bf{k}}}v_{{\bf{k}}}\pi[\delta(\omega+E_{{\bf{k}}})-\delta(\omega-E_{{\bf{k}}})]$. 
Note that chemical potentials $\mu_3$ and $\mu_{\rm S}$ are included in  
$\xi_{{\bf{k}},3/{\rm S}}=\varepsilon_{{\bf{k}}}-\mu_{3/{\rm S}}$, $E_{{\bf{k}}}=\sqrt{\xi_{{\bf{k}},{\rm S}}^2+\Delta_{\rm S}^2}$, and $u_{{\bf{k}}},v_{{\bf{k}}}=\sqrt{(1\pm\xi_{{\bf{k}},{\rm S}} /E_{{\bf{k}}})/2}$.
Inserting these expressions into Eqs.~(\ref{I(1)1})--(\ref{IA}),  
we find that $I^{(1)}$ and $I_1^{(3)}$ vanish, since there is no overlap between $\operatorname{Im}G_{c,11}({\bf{k}},\omega-\Delta\mu)$ and $\operatorname{Im}G_{d,11}({\bf{k}},\omega)$ as long as $\Delta_{\rm S}\neq 0$.  
Similarly, Eq.~(\ref{I2(3)}) is also shown to vanish.
\begin{figure}[tp]
    \centering
    \includegraphics[width=8.6cm]{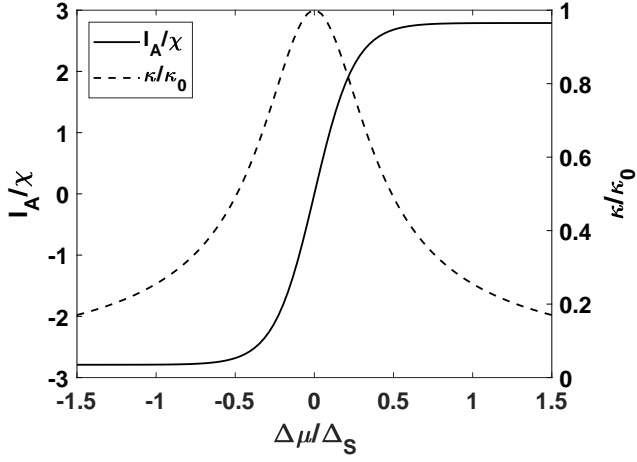}
    \caption{{\bf Tunneling current and conductance.} \red{The solid and dashed lines respectively depict the tunneling current and conductance between two reservoirs as a function of chemical potential bias $\Delta\mu$.} Here $\chi=\Omega^2_{\up,3}\Omega^2_{\dwn,3}mk_{\rm F}/\pi E_{\rm F}^2$ is the normalizing constant for the current, $\kappa_0$ is the constant taken at $\Delta\mu=0$. Values of the chemical potential $\mu_{\rm S}$ and gap energy $\Delta_{\rm S}$ for the superfluid phase are set to be the experimental values in the unitary limit, $\mu_S/E_F=0.38$ and $\Delta_{\rm S}/E_{\rm F}=0.47$ where $E_{F}$, while $T$ is set as $T/T_{\rm F} =0.06$~\cite{PhysRevX.7.041004,PhysRevA.95.043625}\red{($E_{\rm F}$ and $T_{\rm F}$ are the Fermi energy and Fermi temperature of the superfluid, respectively)}.
    }
    \label{IAkappa}
\end{figure}
After inserting $\operatorname{Re}G_{d,12}({\bf{k}},\omega)= -u_{{\bf{k}}}v_{{\bf{k}}}[(\omega-E_{{\bf{k}}})^{-1}-(\omega+E_{{\bf{k}}})^{-1}]$ into Eq.~(\ref{IA}) and performing the momentum integration, we obtain the Andreev current $I_{\rm A}$ as 
\begin{equation}\label{IA1}
    \begin{aligned}
        I_{\rm A}=\Theta(\mu_{\rm S})\Omega^2_{\up,3}\Omega^2_{\dwn,3}
        \frac{m\sqrt{2m\mu_{\rm S}}}{\pi \Delta_{\rm S}^2}
        \frac{e^{\Delta\mu/T}-1}{e^{\Delta\mu/T}+1}.
    \end{aligned}
\end{equation} 
The step function $\Theta(\mu_S)$ in Eq.~(\ref{IA1}) indicates that such tunneling occurs only when the chemical potential of superfluid states is positive.
Notice that all the quantities in Eq.~(\ref{IA1}) can be determined in experiments and hence the result can be directly compared with the experimental result of the nonlinear rf current. This result is valid at weak tunneling coupling where tunneling term is taken as a perturbation, and thus will not be changed qualitatively by higher order corrections when $\Omega_{\sigma,3}$ is small.

If we take $\Delta\mu\rightarrow 0$ with finite temperature $T>0$, $I_{\rm A}$ will reduce to a linear form, $I_{\rm A}=\kappa_0(T)\Delta\mu$, with the conductance $\kappa_0(T)=\Theta(\mu_{\rm S}) \Omega^2_{\up,3}\Omega^2_{\dwn,3}m\sqrt{2m\mu_{\rm S}}/(2\pi\Delta_{\rm S}^2T)$. 
Another interesting fact is that, at zero temperature with finite chemical potential bias, $I_{\rm A}\propto \sign(\Delta\mu)$ does not depend on the magnitude of $\Delta\mu$.
Such a non-Ohmic transport characteristic is nontrivial, since in the conventional N-S interfaces the Andreev currents at a low bias basically obey the Ohm's law even at zero temperature~\cite{PhysRevB.66.165413,PhysRevB.25.4515,PhysRevB.54.7366,PhysRevResearch.2.023340}.
Moreover, $I_{\rm A}$ is not suppressed in the supergap regime ($\Delta\mu>\Delta_{\rm S}$) in contrast to the conventional N-S case.

The tunneling current $I_{\rm A}$ and the conductance $\kappa=I_{\rm A}/\Delta\mu$ between the reservoirs are shown in Fig.~\ref{IAkappa} as functions of $\Delta\mu$. Here $I_{\rm A}$ is normalized by a constant $\chi=\Omega^2_{\up,3}\Omega^2_{\dwn,3}mk_{\rm F}/\pi E_{\rm F}^2$, where $E_{\rm F}=(3\pi^2N_{\rm S})^{2/3}/(2m)$ and $k_{\rm F}=\sqrt{2mE_{\rm F}}$ are respectively the Fermi energy and Fermi momentum for the superfluid, while $\kappa$ is normalized by $\kappa_0$, the conductance at $\Delta\mu=0$. In this figure, we take the values of $\mu_{\rm S}$ and $\Delta_{\rm S}$ as typical experimental values in the unitary limit, $\mu_{\rm S}/E_{\rm F}=0.38$ and $\Delta_{\rm S}/E_{\rm F} =0.47$~\cite{PhysRevX.7.041004,science.1214987,Hoinka2017,PhysRevA.95.043625}, and the temperature is set as $T/T_{\rm F}=0.06$, where $T_{\rm F}$ is the Fermi temperature. We note that the Andreev current in this figure exists not only in subgap, but also in supergap regions.
\begin{figure}[tp]
    \centering
    \includegraphics[width=8.6cm]{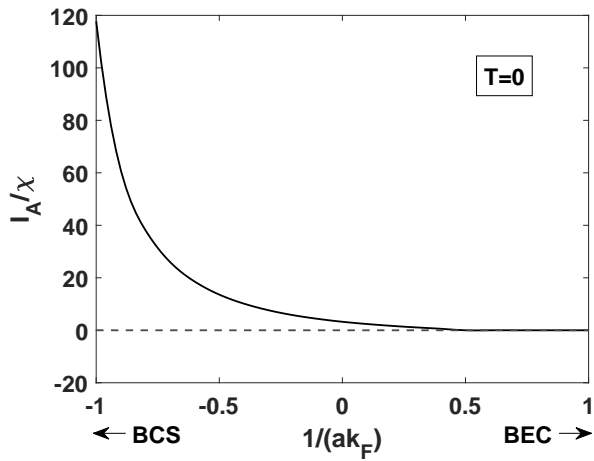}
    \caption{{\bf Andreev current.} The solid line shows the Andreev current as a function of dimensionless coupling $1/(ak_{\rm F})$, where $k_{\rm F}$ is the Fermi momentum and $a$ is the scattering length. We take the limit that $\Delta\mu/T\rightarrow \infty$ for simplicity and $\Delta\mu$ is positive so that the current is on the normal side. Still, $\chi$ is the normalizing constant.}
    \label{I-interaction}
\end{figure}

Moreover, in Fig.~\ref{I-interaction}, we show the tunneling current at zero temperature as a function of dimensionless interaction strength $1/(ak_{\rm F})$, by using the result of $\mu_{\rm S}$ and $\Delta_{\rm S}$ obtained with the diagrammatic approach~\cite{PhysRevA.95.043625,PhysRevResearch.2.023152},
where the scattering length $a$ in the superfluid reservoir is defined by $\frac{m}{4\pi a}=\frac{1}{g}+\frac{m\Lambda}{2\pi^2}$ with the momentum cutoff $\Lambda$~\cite{BCS-BEC}.
We can see that the current decreases monotonically from the BCS limit to the BEC limit as the attraction increases, and becomes zero at around $1/(ak_{\rm F})=0.5$ where $\mu_{\rm S}=0$. This indicates that the Andreev reflection is abundant in the BCS regime, while disappear in the BEC system, even in the momentum-conserved tunneling processes.
This behavior is consistent with that in the previous theoretical work on the spatial N-S junctions without the momentum conservation~\cite{setiawan2021analytic}. Beyond our results that are accurate up to fourth order in $\Omega_{\sigma,3}$, the bosonic Andreev process~\cite{PhysRevLett.102.180405,zapata2011resonant}, in which a pair of incident bosonic particles (holes) is transferred to a pair of bosonic holes (particles), may arise in the BEC limit.
However, the leading order contribution of the bosonic Andreev process 
is proportional to $\Omega_{\sigma,3}^8$~\cite{PhysRevA.106.L011303}, and thus, one can neglect it as far as the analysis up to $\Omega_{\sigma,3}^4$ is concerned.

\begin{figure}[tp]
    \centering
    \includegraphics[width=8.6cm]{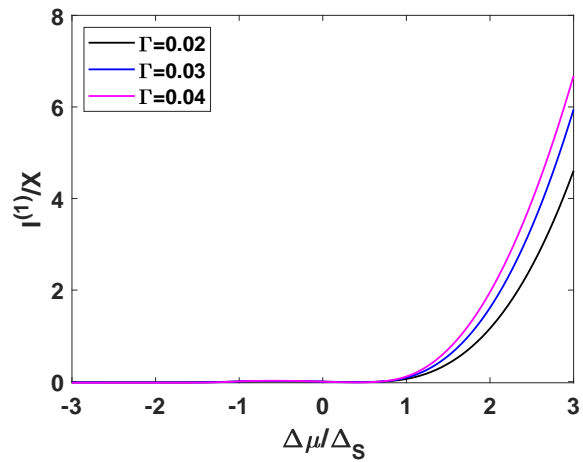}
    \caption{{\bf Quasiparticle-tunneling current with broadened spectral functions.} $X=2mk_{\rm F}(\Omega^2_{\up,3}+\Omega^2_{\dwn,3})/\pi^3$ is the normalization constant for the current, \red{where $k_{\rm F}$ is the Fermi momentum and $\Omega_{\sigma,3}$ represent the Rabi couplings. The black, blue and purple solid lines depict the quasiparticle-tunneling currents with the widths of spectra $\Gamma/E_{\rm F}=0.02$, $0.03$, and $0.04$, respectively.} The values of the chemical potential $\mu_{\rm S}$ and gap energy $\Delta_{\rm S}$ are set to be $\mu_{\rm S}/E_{\rm F}=0.38$ and $\Delta_{\rm S}/E_{\rm F}=0.47$ as those in unitary limit, while $T$ is set as $T/T_{\rm F}=0.06$\red{($E_{\rm F}$ and $T_{\rm F}$ are the Fermi energy and Fermi temperature of the superfluid, respectively)}.}
    \label{fig:I(1)}
\end{figure}

Notice that in our calculation, we have
assumed stable quasiparticles without broadening of the spectral functions, which is valid far away from the critical temperature.
However,
such a broadening can be significant
at finite temperature (in particular, near the critical temperature)
and induce the nonzero contributions of Eqs.~(\ref{I(1)1})-(\ref{I2(3)}) even around the zero detuning. 
To demonstrate broadening effects, we calculate the quasiparticle current with the phenomenological self-energy.
For the lowest-order quasiparitcle current $I^{(1)}$, its dependence on the broadening effect is shown in Fig.~\ref{fig:I(1)}, where the imaginary part of self-energy is introduced into Green's functions as $\Sigma_{11}=-\Sigma_{22}=-i\Gamma$. We find that for positive $\Delta\mu$, $I^{(1)}$ is suppressed in subgap region and shows a nearly Ohmic transport in supergap region. However, since the overlap between two spectra becomes small and no Fermi surface exists in the normal phase for large negative $\Delta\mu$, the quasiparticle current is suppressed for negative $\Delta\mu$ even in supergap region. 
In BCS region with small $\Delta_{\rm S}$, the spectral broadening smears out the excitation gap around the zero detuning, leaving a strong spectra of quasiparticle tunneling~\cite{doi:10.1126/science.1100818}.
Nevertheless, by considering the crossover regime with moderate gap sizes~\cite{PhysRevLett.101.140403} at $T/T_{\rm F}\ll 1$, we can sufficiently suppress these quasiparticle tunneling processes and distinguish the signal of $I_{\rm A}$ at zero detuning. 

If we consider nonzero detunings, namely, out of rf resonances, a shift will be added on the chemical potential bias $\Delta\mu$~\cite{PhysRevA.64.033609}. As a result, the Green’s functions in Eqs.~(\ref{I(1)1})--(\ref{IA}) which include $\Delta\mu$ will also undergo shifts on the energy dependence. We find that with a shift on $\Delta\mu$, the quasiparticle tunneling $I^{(1)}$ and $I^{(3)}_1$, and pair tunneling processes $I_2^{(3)}$ will occur due to the overlap between spectral functions (Dirac functions) even at zero temperature. On the other hand, the nonzero detuning will only give a shift on $\Delta\mu$ in the formula of Andreev current $I_{\rm A}$ Eq.~(\ref{IA1}): $I_{\rm A}(\mu_{\rm S},\Delta\mu)\rightarrow I_{\rm A}(\mu_{\rm S},\Delta\mu+\delta)$ with $\delta$ denoting the detuning. In this regard, the Andreev current, quasiparticle current, and pair tunneling current can coexist out of rf reonances. One way to distinguish the Andreev current is to tune the temperature above the superfluid critical temperature $T_{\rm c}$, where all components are in normal phase and the Andreev current disappears. By comparing the signal below $T_{\rm c}$ and the one above $T_{\rm c}$, the Andreev current can be extracted from the total signal. 
To do this, the broadened spectral functions at finite temperature should be taken into account accurately, which is left for future work.

Since two reservoirs are spatially overlapped in our synthetic N-S junction, there exist residual interactions between normal and superfluid components, apart from the strong interaction within the superfluid. This will cause self-energy shifts in each reservoir. A well-known correction in the weakly interacting limit is the Hartree shift \cite{PhysRevA.85.012701}, given by $\Sigma_{\sigma}=\frac{4\pi a_{\sigma 3}}{m}N_{3}$ and $\Sigma_{3}=\sum_{\sigma}\frac{4\pi a_{\sigma 3}}{m}N_{\sigma}$, where $a_{\sigma 3}$ is the scattering length between components $\ket{\sigma}$ and $\ket{3}$. 
These give an effective shift of the chemical potentials as $\mu_3^{\rm eff}=\mu_3-\Sigma_{3}$ and $\mu_{\sigma}^{\rm eff}=\mu_\sigma-\Sigma_{\sigma}$. Consequently, the chemical potential bias will be modified as
    $\Delta\mu^{\rm eff}=\mu_3^{\rm eff}-\mu_{\rm S}^{\rm eff}=\Delta\mu-\Sigma_{3}+\frac{\Sigma_{\uparrow}+\Sigma_{\downarrow}}{2}$,
where $\mu_{\rm S}^{\rm eff}=(\mu_{\up}^{\rm eff}+\mu_{\dwn}^{\rm eff})/2$ is the averaged effective chemical potential in the superfluid. Therefore, by replacing $\Delta\mu$ with $\Delta\mu^{\rm eff}$ in Eq.~(\ref{IA1}), the result is adapted to the case with weak residual interactions between reservoirs. On the other hand, an effective magnetic field $h_{\rm eff}=(\Sigma_{\up}-\Sigma_{\dwn})/2$ arises in the superfluid due to the unbalanced residual interactions, even in the balanced mixture $\mu_{\up}=\mu_{\dwn}$. However, it can be negligible if $h^{\rm eff}$ is sufficiently small compared to $\Delta_{\rm S}$. 

In the case of the $^6$Li three-component mixture, it is known that the strong three-body loss shortens the system's lifetime~\cite{PhysRevLett.101.203202,PhysRevLett.102.165302}.
The timescale of tunneling processes can be estimated by the uncertainty principle: $\Omega \tau\geq\frac{1}{2\pi}$ (where we take $\Omega_{3,\up}=\Omega_{3,\dwn}\equiv\Omega$ for simplicity).
Taking the typical magnitude of Fermi energy in $^6{\rm Li}$ Fermi gases as $10^3$ Hz and $\Omega/E_{\rm F}\sim 0.1$ to justify the perturbative treatment, we have $\tau\sim 10^{-3}$s, which is smaller than the timescale of atom losses. 
More precisely, the particle transport timescale can be estimated by using $\tau'=\beta \kappa^{-1}$~\cite{brantut2013thermoelectric}, where $\beta=\frac{\partial N}{\partial\mu}|_T$ is the compressibility of the reservoirs and $\kappa$ is the conductance of the particle current. The dimension of $\beta$ is given by $\beta\sim \frac{mk_{\rm F}}{2\pi^2}$. Then according to the expression of the conductance $\kappa_0$ of the Andreev current, we have $\tau'\sim\Delta^2_{\rm S} T/(\pi\Omega^4)$. In this sense, by adjusting the value of Rabi frequencies $\Omega$, one can tune the transport timescale to be smaller than the timescale of atom losses. In either way, this indicates that one can measure the Andreev current before encountering the significant particle-number losses.

A promising way to detect the Andreev current with avoiding the three-body loss would be the preparation of the solely superfluid state with two-component fermions before applying the Rabi coupling.
In this case, the normal component is dilute and therefore largely negative $\mu_3$ (i.e., largely negative $\Delta\mu$) would be realized.
As we showed in Fig.~\ref{IAkappa},
still one may find the nonzero Andreev current in such a supergap regime.
Moreover, in order to avoid the the overlap of Feshbach resonances leading to the strong three-body losses, it is possible to use higher hyperfine states as the normal component~\cite{Ketterle2008}.

\subsection*{Summary}
In this work, we investigate the particle tunneling through an effective N-S interface designed by two rf laser fields that hold the momentum conservation. 
By addressing the nonlinear response regime in terms of the Schwinger-Keldysh formalism, we
find that the Andreev reflection is the only process passing through the synthetic interface up to the fourth-order perturbation in $H_{\rm t}$.
We succeed in obtaining the analytical solution of the current and
show the dependence of Andreev current and conductance on the chemical bias between two reservoirs. 
We also demonstrate how the Andreev current at zero temperature varies with the interaction strength, from the BCS to BEC regime. Another interesting outcome is that, different from conventional cases, the present tunneling current totally violates Ohm's law at zero temperature.
  
\section*{Discussion}
Our proposed system inducing the momentum-conserved tunneling may also be promising for understanding the black hole information paradox.
Some similarities can be found between the momentum-conserved Andreev reflection and Hayden-Preskill model~\cite{Hayden_2007,Lloyd2014UnitarityOB}, where certain final states in black hole allow the teleportation of information contained in matter falling into the black hole to the Hawking radiation going outwards~\cite{Hawking1975,Horowitz_2004}.
 In our case, the BCS superfluid can be regarded as the black hole final states, which permit transferring of the quantum information encoded in an incident particle (hole) from the normal side to an outgoing hole (particle). Since the tunneling process is momentum-conserved, the hole is reflected with exactly the opposite momentum to the incident particle (hole), which ensures that the same information is teleported to the reflected one.

To further study this topic, we may prepare two hyperfine states in the normal side, for example, $\ket{3}$ and $\ket{4}$ with two Rabi couplings: $\Omega_{\up,3}$ and
$\Omega_{\dwn,4}$. In this case, we can consider an incident particle at a superposition state of states $\ket{3}$ and $\ket{4}$, and investigate the reflected mode, which is expected to be in the same superposition state as the incident one. 
To be self-contained, we discuss how the information mirror process proposed in Ref.~\cite{PhysRevD.96.124011} can be realized in this system.
In the following, $\ket{1}_{{\bf{k}},3}\equiv c_{{\bf{k}},3}^\dag\ket{0}$ and $\ket{1}_{{\bf{k}},4}\equiv c_{{\bf{k}},4}^\dag\ket{0}$ are regarded as $\ket{1}_{{\bf{k}},\up}\equiv c_{{\bf{k}},\up}^\dag\ket{0}$ and $\ket{1}_{{\bf{k}},\dwn}\equiv c_{{\bf{k}},\dwn}^\dag\ket{0}$ in the normal side, respectively. 
We consider an incident mode from the normal phase $\ket{\phi_c}=(ac^\dagger_{{\bf{k}},3}+bc^\dagger_{{\bf{k}},4})\ket{0}$ and a tunneling Hamiltonian $H'_t=\Omega\sum_{\sigma}(d^\dagger_{{\bf{k}},\sigma}c_{{\bf{k}},\sigma}+{\rm H.c.})$ where $a,b\in \mathbb{C}$ and $\Omega_{3,\uparrow}=\Omega_{4,\downarrow}\equiv \Omega$ is taken for simplicity.
A combined state in the direct product space of a single incident particle and a particle-hole pair at the interface is defined as $\ket{\psi}=\ket{\phi_c}\otimes(d^\dagger_{{\bf{q}},\up}c_{{\bf{q}},\up}+d^\dagger_{{\bf{q}},\dwn}c_{{\bf{q}},\dwn})\ket{G}$, where $\ket{G}=\ket{1_{{\bf{q}},\up}1_{{\bf{q}},\dwn}\cdots}$ denotes the fulfilled Fermi sea. The incident mode evolving with the tunneling Hamiltonian reads
\begin{equation}
    \ket{\phi(\tau)}=e^{-i\tau H'_{\rm t}} \ket{\phi_c}=\sin{(\Omega\tau)}\ket{\phi_d}+\cos{(\Omega\tau)\ket{\phi_c}},
\end{equation}
where
$\ket{\phi_d}=(ad^\dagger_{{\bf{k}},\up}+bd^\dagger_{{\bf{k}},\dwn})\ket{0}$, and $\tau$ is the smallest time interval for the quantum system to make a change. While satisfying the uncertainty principle, $\tau\simeq\pi/2\Omega$ is adopted to permit a complete mode transfer, $c^\dagger_{{\bf{k}},\sigma}\rightarrow d^\dagger_{{\bf{k}},\sigma}$. The combined state then becomes $\ket{\psi}\rightarrow\ket{\phi_d}\otimes(d^\dagger_{{\bf{q}},\up}c_{{\bf{q}},\up}+d^\dagger_{{\bf{q}},\dwn}c_{{\bf{q}},\dwn})\ket{G}$. The BCS ground state $\ket{\Psi_{\rm BCS}}=\prod_{{\bf{k}}}(u_{{\bf{k}}}+v_{{\bf{k}}}d^\dagger_{{\bf{k}},\up}d^\dagger_{-{\bf{k}},\dwn})\ket{0}$ treated as the final state is imposed on the combined state, which yields a reflected state,
\begin{equation}
    \braket{\Psi_{\rm BCS}}{\psi}\propto\ket{\phi_h}.
\end{equation}
We note the reflected mode $\ket{\phi_h}=(a h_{-{\bf q} \up}^{\dagger}+b h_{-{\bf q} \dwn}^{\dagger})\ket{0'}$, where $\ket{0'}$ denotes the quasiparticle vacuum while $h^\dagger_{-{\bf{q}},\dwn}\ket{0'} =\ket{0_{{\bf{q}},\up}1_{{\bf q},\dwn}\cdots}$ and $h^\dagger_{-{\bf{q}},\up}\ket{0'}=\ket{1_{{\bf{q}},\up}0_{{\bf q},\dwn}\cdots}$ denote holes, is a hole-like mode in the same spin state as the incident mode. This indicates that the quantum information is transferred from the incident mode to the reflected one, which is known as the deterministic teleportation.

\section*{Methods}
We apply the Schwinger-Keldysh Green's function formalism to calculate the tunneling current in a non-equilibrium steady state. We use the expanded Keldysh contour, which includes two parts along the real time axis: a forward contour (from $t=-\infty$ to $t=\infty$) and a backward contour (from $t=\infty$ to $t=-\infty$). The current can be expressed in terms of lesser Green's functions
\begin{equation}
    \begin{aligned}
        G^<_c({\bf{k}}, t_-,t'_+)&=i\langle c^\dagger_{{\bf{k}}}(t'_+)c_{{\bf{k}}}(t_-)\rangle,\\
        G^<_d({\bf{k}}, t_-,t'_+)&=i\langle d^\dagger_{{\bf{k}}}(t'_+)d_{{\bf{k}}}(t_-)\rangle,
    \end{aligned}
\end{equation}
where $t_-$ and $t'_+$ respectively denote the time arguments on the forward and backward parts. The integral over the Keldysh contour can be changed into that over the real time axis according to the Langreth rules, which read 
\begin{equation}
    C(t,t')=\int_C dt_1\,A(t,t_1)B(t_1,t'),
\end{equation}
\begin{equation}
    \begin{aligned}
        C^{\gtrless}(t,t')=&\int_{-\infty}^{\infty} dt_1\,[A^{\rm ret.}(t,t_1)B^{\gtrless}(t_1,t')\\
        &+A^{\gtrless}(t,t_1)B^{\rm adv.}(t_1,t')],
    \end{aligned}
\end{equation}
\begin{equation}
    C^{\rm ret.}(t,t')=\int_{-\infty}^{\infty} dt_1\, A^{\rm ret.}(t,t_1) B^{\rm ret.}(t_1,t').
\end{equation}
Here $A$, $B$, and $C$ denote arbitrary time-ordering correlation functions and the superscripts ``$>$'' and ``adv.'' are respectively for greater and advanced correlation functions. Since the Green's function $G^{\rm ret.}(t,t')$ or $G^<(t,t')$ only depends on the time difference $t-t'$, its Fourier transform depends on a single frequency $\omega$. Therefore, we obtain Eq.~(\ref{I(1)}) for the lowest order term of the momentum-conserved tunneling current. The expressions for higer order terms, $I^{(3)}_1$, $I^{(3)}_2$, and $I_{\rm A}$ are obtained similarly. 

\section*{Data Availability}
\red{Data supporting the findings of this study are available from the corresponding author upon reasonable request.}

\section*{Code Availability}
\red{The code used for the numerical calculations in this study are available from the corresponding author upon reasonable request.}

\section*{References}

\bibliographystyle{sn-nature.bst}
\bibliography{ref.bib}

\begin{acknowledgements}
The authors thank RIKEN iTHEMS NEW working group and M. Horikoshi for fruitful discussions.
HT is supported by JSPS KAKENHI under Grant Nos.~18H05406 and 22K13981.
YS is supported by JSPS KAKENHI under Grant No.~19J01006.
SU is supported by MEXT Leading Initiative for Excellent Young Researchers, JSPS KAKENHI under Grant No.~21K03436, and Matsuo Foundation.
HL is supported by JSPS KAKENHI under Grants Nos.~18K13549 and 20H05648.
YS and HL are supported by the RIKEN Pioneering Project: Evolution of Matter in the Universe (r-EMU).
\end{acknowledgements}

\section*{Author contributions.}
T. Z. carried out the study.
T. Z. wrote the first draft and H. T., Y. S., S. U., and H. L. edited the manuscript.
All the authors discussed the results and reviewed the manuscript.

\section*{Competing interests.}
The authors declare no competing interests.

\end{document}